\documentclass[aps,prd,preprint,superscriptaddress,showpacs,nofootinbib]{revtex4}
\usepackage{graphics}
\usepackage{epsfig}
\usepackage{latexsym}
\usepackage{colordvi}
\usepackage{amsmath}
\usepackage{amssymb}

\newcommand{\be}{\begin{equation}}
\newcommand{\ee}{\end{equation}}
\newcommand{\bear}{\begin{eqnarray}}
\newcommand{\eear}{\end{eqnarray}}
\newcommand{\ba}{\begin{array}}
\newcommand{\ea}{\end{array}}

\begin{document}
\preprint{\parbox[b]{1in}{ \hbox{\tt PNUTP-08/A01,\hskip 0.1in
KIAS-P08029 } {\tt IC/2008/012} }}

\title{Holographic Monopole Catalysis of Baryon Decay}

\author{Deog Ki Hong}
\email[E-mail: ]{dkhong@pusan.ac.kr} \affiliation{Department of
Physics, Pusan National University,
             Busan 609-735, Korea}

\author{Ki-Myeong Lee}
\email[E-mail: ]{klee@kias.re.kr} \affiliation{School of Physics,
Korea Institute for Advanced Study, Seoul 130-722, Korea}

\author{Cheonsoo Park}
\email[E-mail: ]{chspark@pusan.ac.kr} \affiliation{Department of
Physics, Pusan National University,
             Busan 609-735, Korea}

\author{Ho-Ung Yee}
\email[E-mail: ]{hyee@ictp.it} %\\
\affiliation{The Abdus Salam International Centre for
Theoretical Physics, Strada Costiera 11, 34014, Trieste, Italy} \vspace{0.1in}

\vspace{0.1in}

\date{\today}

\begin{abstract}
We study how monopole catalysis of baryon decay is realized in
holographic QCD. Physics of monopole catalysis becomes much simpler in
holographic description as it occurs due to the violation of the
Bianchi identity for the 5D gauge symmetry when magnetic monopole is
present. In holographic QCD we find a unified picture of the baryon
number violation under magnetic monopole or electroweak sphaleron,
giving a new mechanism of baryon number violation.
We also embed our set-up in the string theory model by Sakai and Sugimoto.

\end{abstract}
%%% 12.38.-t Quantum chromodynamics
%%% 11.25.Tq Gauge/string duality
%%% 12.38.Aw General properties of QCD
%%% 14.20.Dh protons and neutrons
%%% 11.10.Kk Field theory in dimension other than four

\maketitle
\newpage
\section{Introduction}
Recent development~\cite{Maldacena:1997re} of the dual correspondence
between gauge theories and string theories
has given us a powerful tool to investigate
strongly coupled gauge theories like Quantum Chromodynamics (QCD) or Technicolor, which are otherwise
extremely difficult to solve.
According to the gauge/string duality the low energy QCD becomes a theory of mesons in
the large number of color ($N_c$)
and large 't Hooft coupling ($\lambda\equiv g_s^2N_c$) limit, but in a warped five-dimensional spacetime.
The theory, known as holographic QCD, is nothing but a 5D flavor gauge theory in the
warped background geometry, endowed with a Chern-Simons term,
necessary to realize the global anomalies of QCD~\cite{Sakai:2004cn,Erlich:2005qh}.

Being the theory of mesons in the large $N_c$ limit, holographic QCD
should admit baryons as solition solutions, as conjectured by Skyrme
long time ago for the nonlinear sigma model~\cite{Skyrme:1961vq}.
Indeed, it was found that the baryons are realized as instanton
solitons in holographic QCD \cite{Son:2003et}\footnote{There is an alternative realization of baryons
through 5D holographic baryon fields \cite{deTeramond:2005su}.}.
The instanton picture of baryons
reproduces the success of skyrmions rather well but with much less
parameters for the spectrum and the static properties of
baryons~\cite{Hong:2007kx,Hata:2007mb,Nawa:2006gv,Hong:2007dq}.
Unlike skyrmions, however, the instanton solitons are made of not
only pions but infinite towers of vector mesons, intertwined
nontrivially, leading to small size objects without any intrinsic
core, which therefore realizes full vector meson dominance for
baryons~\cite{Hong:2007kx,Hong:2007dq}.

One of nice features of skyrmion is that monopole catalysis of
baryon decay~\cite{Rubakov:1981rg,Callan:1982ac} is easily described
in the skyrmion picture of baryons, which then sets up the practical
basis for the calculation of monopole
catalysis~\cite{Callan:1983nx}. The magnetic monopole provides a
defect on which the topological charge of skyrmions can unwind,
allowing skyrmions to decay at the rate of QCD scale without
suppression by the monopole scale.

In this paper we describe how monopole catalysis of baryon decay
realizes in holographic QCD. The magnetic monopole catalyzes baryon
decay, since the 5D baryon number current,
$B^{M}=(1/32\pi^2)\epsilon^{MNPQR} {\rm Tr}\,F_{NP}F_{QR}$ is not
conserved in the presence of the holographic magnetic monopole
string by violating the Bianchi identity for the gauge fields.
Furthermore, we show that monopole catalysis can be naturally
described in string theory as the dissolution of $D4$ brane
(instanton soliton) into $D6$ brane (monopole string), which
suggests that  monopole catalysis of baryon decay should occur
without any barrier and hold even beyond the large $N_c$ limit. The
decay rate of baryons by monopole catalysis is determined by the
scale of instanton solitons. In section 2 we first briefly review
the monopole catalysis of skyrmion decay, studied by Callan and
Witten~\cite{Callan:1983nx}, and then in sections 3 and 4 we
describe how monopole catalysis is realized in holographic QCD.
Finally in section 5 we present the string theory realization of
monopole catalysis of baryon decay. Section 6 contains concluding
remarks together with future directions.

\section{Monopole Catalysis of Baryon Decay\label{section2}}

Monopoles can arise in grand unified theories (GUTs) as the unified gauge symmetry
breaks down to the Standard Model gauge group. Their typical size is
about the unification scale, which is nearly point-like compared to
the usual scales of the Standard model physics, especially QCD.
Since electromagnetism is the only long-ranged interaction in the
Standard Model, a GUT monopole eventually looks like an
electromagnetic Dirac monopole to low energy observers. Near the
core of the monopole, there are clouds of heavy GUT gauge fields
which can mediate various GUT interactions, among which are baryon
number violating processes. Despite its small size, the cross
sections for these monopole-induced processes are not suppressed by
the unification scale, due to their origin being related to chiral
anomalies. Rather, the monopole center provides a baryon number
violating vertex of unit strength acting as a catalysis of baryon
decay, and the actual cross section is governed by low energy
dynamics such as QCD.

The physics gets more interesting when the low energy dynamics such
as QCD becomes strongly coupled and takes a completely different
looking effective theory. Because the monopole catalysis is not
suppressed by any scales of the UV theory, its presence must persist
even in the low energy effective theory, and we need a
 non-trivial disguise of the monopole-induced baryon
decay as an interesting low energy phenomenon within the effective
theory. As the structure of the monopole core given by the UV
physics is not a concern of the low energy effective theory, the
Dirac monopole profile of the unbroken gauge group should be
sufficient for the low energy description of monopole catalysis.
This indicates an intricate theoretical consistency requirement for
the physics induced by Dirac monopoles in any low energy effective
theory.

A consistent low energy effective theory of QCD is the chiral
Lagrangian of $SU(N_F)_L\times SU(N_F)_R$, and the low energy
baryons are effectively described by topological solitons, called
Skyrmions, in the large $N_c$ limit. Assuming that the large $N_c$ QCD can
be embedded in some GUT theory which contains monopoles capable of
baryon number catalysis, the above discussion leads to the
expectation that an electromagnetic Dirac monopole should  be able
to induce Skyrmion-baryon decay within the low energy chiral
dynamics. At first sight, this looks puzzling because the baryon
number of the Skyrmions is purely topological and there seems to be
no way for the Skyrmions to decay within the framework of the
effective theory. This is the problem of monopole catalysis of
Skyrmion decay analyzed by Callan and Witten long ago~\cite{Callan:1983nx}.

The resolution of the puzzle is that the baryon number and its
current must be modified in the presence of a background
electromagnetic field~\footnote{In general, any background
$SU(N_F)_L\times SU(N_F)_R\times U(1)_B$ gauge fields upon weakly
gauging it requires modification of the baryon number current.}, to
be gauge invariant and conserved at the same time. This requirement
determines the baryon number current uniquely~\cite{Callan:1983nx}.
More explicitly, the standard baryon number current of Skyrmions
which is topologically conserved is given by
\be
B^\mu={1\over
24\pi^2}\epsilon^{\mu\nu\alpha\beta}{\rm Tr}\left(U^{-1}\partial_\nu
U U^{-1}\partial_\alpha U U^{-1}\partial_\beta U \right)\quad,\label{skyrmion}
\ee
where $U$ is the $SU(2)$ group field of the chiral
Lagrangian~\footnote{For simplicity, we will confine our discussion
to the $N_F=2$ massless quarks} which transforms as
\be
U \to g_L U
g_R^\dagger\quad,
\ee
under the chiral symmetry $SU(2)_L\times
SU(2)_R$. Since the electromagnetic $U(1)_{EM}$ acts on $U$ by
\be
U\to e^{iQ}U e^{-iQ}\quad,\quad Q=\left(\begin{array}{cc}{2\over 3}
& 0\\0 & -{1\over 3}\end{array}\right)\quad,
\ee
the simplest way to
make the baryon current gauge invariant would be to replace the
ordinary derivatives $\partial U$ by the covariant derivatives
$DU=\partial U +A^{EM}[Q,U]$, where $A^{EM}$ is the electromagnetic
gauge potential~\footnote{We use the convention where the gauge
potential is {\it anti}-hermitian, and the covariant derivative is
$D=\partial+A$. To compare with Ref.\cite{Callan:1983nx}, simply
replace $A$ by $-ieA$.}. However, the resulting baryon current is no longer
conserved in general, and we need to add more gauge-invariant
terms to make it conserved. This has been worked out in
Ref.\cite{Callan:1983nx}, and we quote the result
\bear B^\mu &=&
{1\over
24\pi^2}\epsilon^{\mu\nu\alpha\beta}{\rm Tr}\left(U^{-1}\partial_\nu
U U^{-1}\partial_\alpha U U^{-1}\partial_\beta U \right)\nonumber\\
&& -{1\over
24\pi^2}\epsilon^{\mu\nu\alpha\beta}\partial_\nu\left[3A^{EM}_\alpha
{\rm Tr}\left(Q(U^{-1}\partial_\beta U+\partial_\beta U U^{-1})\right)\right]\quad.\label{newbn}
\eear
It is clear that the final form of the
modification does not affect the conservation of the current, while it can also be checked that
the result is $U(1)_{EM}$ gauge invariant.

However, the conservation of baryon number,
$\partial_\mu B^\mu=0$, is guaranteed only for a smooth, well-defined
background potential $A^{EM}$. Since a Dirac monopole field is
singular and doesn't have a well-defined potential, its presence
might invalidate the conservation of the above gauge invariant
baryon number. Indeed, this is exactly what
causes Skyrmions to decay in the presence of a monopole.
One might question that the topological Skyrmion number (\ref{skyrmion})
can never be violated under a smooth time evolution, and an initial
Skyrmion would never decay to topologically trivial meson states.
However, a caveat is that a Dirac monopole entails the Dirac-string on which
$A^{EM}$ is singular, and
we are allowed to take singular gauge transformations to move this string from
one direction to another. Under these singular $U(1)_{EM}$ gauge transformations which are
now allowed
in a Dirac monopole background,
the topological Skyrmion number can actually change. In other words, the Skyrmion number
is not a well-defined gauge-invariant quantity in the presence of a Dirac monopole, and
a configuration with non-zero Skyrmion number can be
equivalently described by a topologically trivial configuration under a gauge transformation.
What remains invariant under the gauge transformations is the new gauge-invariant
baryon number (\ref{newbn}).

This gauge-invariant baryon number can be eaten up at the monopole center dynamically,
which is responsible for the baryon decay.
Near the center of the monopole, only the neutral component of the pion, $\pi^0$,
can take non-zero values, while charged pion excitations would cause too much energy, since they have nonzero
angular momentum proportional to the magnetic charge of monopole.
Writing $U(t)={\rm exp}({2i\over F_\pi} \pi^0(t) \sigma^3)$ near the center
of a Dirac monopole of unit strength
\be
A^{EM}=-{i\over 2}(1-\cos\theta)d\phi\quad,
\ee
and using $\epsilon^{r\theta\phi t}={-1\over \sqrt{g}}={-1\over r^2\sin\theta}$ in the polar coordinate,
the radial flux of the baryon number out of the monopole is readily calculated to be
\be
B^r={\partial_t\pi^0\over 4\pi^2 F_\pi r^2}\quad,
\ee
whose integration gives the change of baryon number
\be
{dB\over dt}={1\over \pi F_\pi  }({\partial_t\pi^0})\quad.\label{change}
\ee
Therefore, the rate of change of $\pi^0$ at the monopole center
is proportional to disappearance of the baryon charge from the effective theory.
In the original GUT, this baryon number violation should be accompanied by
creation of leptons, whose detail should resort to some unknown
dynamics at the center of the monopole, so that the total fermion number is conserved.
In the low energy effective theory,
this normally involves putting a relevant boundary condition
on the leptons at the monopole center, such that the change of baryon number is compensated by
the change of lepton number.
In the following sections, we will see that all the above features nicely fit into
a simple description in the framework of holographic QCD.

\section{Holographic Baryon Number Current}

To be specific, we present our analysis in the model by Sakai and Sugimoto (SS) from Type IIA
string theory (See Ref.\cite{Bergman:2007pm,Casero:2007ae,Aharony:2008an} for its quark mass deformation).
However, most of the steps we perform are dictated by symmetry and don't in fact
depend on the details of the model, and hence
our analysis is applicable to any model of holographic QCD.

In the SS model, the world volume $U(N_F)_L$ and $U(N_F)_R$ gauge fields
on $D8$ and $\bar D8$ in UV region are holographically dual to
the corresponding chiral symmetry in QCD. Its spontaneous breaking
to the diagonal $U(N_F)_I$ is geometrically realized
by adjoining $D8$ and $\bar D8$ at the tip of the cigar geometry.
Alternatively, we can view this as having a single $D8$ brane whose
two asymptotic boundaries towards UV region encode the chiral symmetry
$U(N_F)_L$ and $U(N_F)_R$ respectively.
The latter view point is more practical in the analysis, and
we introduce a coordinate $z$ on $D8$ such that $z\to\pm \infty$
represent two UV boundaries. We will call $z$ the radial or the 5th direction.
Assuming homogeneity along the internal $S^4$ fibration,
the world volume theory on $N_F$ $D8$ branes is effectively a 5D $U(N_F)$ gauge
theory in a non-trivial $z$ dependent background.
According to AdS/CFT correspondence, the asymptotic values
of the 5D gauge potential near the two boundaries, $A_\mu(x,z\to\infty)$
and $A_\mu(x,z\to-\infty)$\footnote{We use Greek indices for the Minkowki $R^{1,3}$ directions,
while the full 5D coordinates $(x^\mu,z)$ will be denoted by capital letters.},
are non-dynamical background fields coupled to QCD $U(N_F)_L$ and $U(N_F)_R$ currents respectively.
Equivalently, they are
precisely the background gauge potential upon weakly gauging the chiral symmetry.

Note that the above prescription holds true in the gauge where $A_z$
is kept free (and vanishes at $z\to\pm\infty$), while it is often more convenient
to work in the gauge where $A_z=0$. After performing a suitable
gauge transformation from the above to the $A_z=0$ gauge, the boundary behavior of $A_\mu$
will be slightly different from the above. However, any gauge invariant calculations
are independent of the gauge choice.

The 5D gauge theory on $D8$ also contains a tower of normalizable (axial) vector meson excitations
in view of 4D observers. Especially, the Wilson line
\be
U(x^\mu)=P \exp \left( -\int_{-\infty}^{+\infty}dz \, A_z (x^\mu,z)\right)\quad,\label{wilson}
\ee
is identified as the massless Nambu-Goldstone pion for the chiral symmetry breaking
$U(N_F)_L\times U(N_F)_R\to U(N_F)_I$~\footnote{Note that axial anomaly of $U(1)_A$
is negligible in the large $N_c$ limit.}, and it is precisely the group field entering
the low energy QCD chiral Lagrangian.
Upon expanding $A_M$ in terms of normalizable (axial) vector mesons as well as
the non-normalizable background fields previously mentioned~\footnote{The modes from $A_z$,
except the Wilson line (\ref{wilson}),
are eaten after all
 by massive spin 1 (axial) vector mesons coming from $A_\mu$. In this sense, the $A_z=0$
gauge is a kind of unitary gauge where only physical degrees of freedom are present.},
and performing
$z$-integration we obtain a 4D effective chiral Lagrangian of $U(x)$ and
excited mesons coupled to the background gauge potential. The part that contains $U(x)$
reproduces the previously known gauged Skyrmion theory with the correct Wess-Zumino-Witten term.
Therefore, the 5D gauge field compactly summarizes the pions, the excited mesons,
and the background gauge potential of the chiral symmetry in a single unified framework.

As the 5D gauge theory on $D8$ branes includes the Skyrmion theory,
there must exist topological objects similar to Skyrmions that play a role of baryons.
Indeed, the 5D gauge theory has topological solitons whose field profile
on the spatial $(x^\mu, z)$ directions taking that of instantons.
We call them instanton-baryons not to be confused with real instantons in Euclidean field theory.
This also agrees with the baryonic objects coming from $S^4$-wrapped $D4$ branes in the string theory
of this background, since these $D4$ branes can dissolve into $D8$ branes exactly as instanton-solitons~\cite{Douglas:1995bn}.
The topology of instanton-baryons is counted by the instanton number
\be
B={1\over 32 \pi^2}\int dz dx^3 \epsilon^{MNPQ}{\rm Tr}\left(F_{MN} F_{PQ}\right)
={1\over 8\pi^2}\int_{R^4}{\rm Tr}\left( F\wedge F\right)\quad,\label{number}
\ee
where $M,N,P,Q$ spans only spatial dimensions and the epsilon tensor is defined in the flat space.
In our convention, $F=dA+A\wedge A$.
Being topological, its conservation is guaranteed in any smooth situations.

We can easily find the corresponding conserved current in 5D \cite{Son:2003et}.
From the Bianchi identity $DF\equiv dF+A\wedge F-F\wedge A=0$,
we have
\be
d{\rm Tr}\left( F\wedge F\right)={\rm Tr}\left(DF\wedge F\right)+{\rm Tr}\left(DF\wedge F\right)=0\quad,
\ee
and the current 1-form defined by $j_B=*_5{\rm Tr}\left(F\wedge F\right)$ is conserved
\be
d*_5 j_B=0\quad.\label{conserve}
\ee
Note that this is independent of what metric we use in defining $*_5$ since the conservation
is a consequence of the Bianchi identity. In fact, the results of the following discussions
will not be affected by metric at all,
and for simplicity
we will keep the flat 5D metric in $(x^\mu,z)$ coordinate whenever we need the metric as
intermediate steps.
In components (\ref{conserve}) is written as $D_M j_B^M=\partial_M j_B^M=0$,
where $j_B^M=g^{MN} j_{BN}=\eta^{MN} j_{BN}$ and $D_M$
is the metric covariant derivative.
This means that we can define a 4D conserved current $B^\mu$ by simply integrating
$j_B^\mu$ along the $z$-direction,
\be
B^\mu\equiv \int_{-\infty}^{+\infty} dz \, j_B^\mu\quad,
\ee
where the conservation is shown by
\be
\partial_\mu B^\mu =\int_{-\infty}^{+\infty} dz \, \partial_\mu j_B^\mu
=-\int_{-\infty}^{+\infty} dz \, \partial_5 j_B^5=j_B^5(-\infty)-j_B^5(+\infty)=0\quad,\label{consv}
\ee
as the boundary term
\be
j_B^5(\pm\infty) \sim \epsilon^{\mu\nu\alpha\beta}{\rm Tr}
\left(F_{\mu\nu}F_{\alpha\beta}\right)|_{z\to\pm\infty}\label{bdr}
\ee
is the chiral sphaleron density of the background gauge potential of the chiral symmetry, and
we assume that it vanishes for now. This is justified in our consideration of
static monopoles without electric fields which will be discussed in a moment.
We will come back to the implication of these boundary terms later.
The explicit form of the 4D conserved baryon current $B^\mu$ is
\be
B^\mu=\int_{-\infty}^{+\infty}dz \,j_B^\mu
\sim\int_{-\infty}^{+\infty}dz \,\epsilon^{\mu\nu\alpha\beta }\,
{\rm Tr}\left(F\wedge F\right)_{\nu\alpha\beta z}=\eta^{\mu\nu}\left[
*_4 \int_z\,{\rm Tr}\left(F\wedge F\right)\right]_\nu\quad.
\ee
The final form makes it clear that the end result is indeed metric-independent.
By comparing with the normalized instanton number (\ref{number}) as $\int d^3x B^0$,
we can easily fix the normalization
to be
\be
B^\mu={1\over 8\pi^2}\int_{-\infty}^{+\infty} dz\,\epsilon^{\mu\nu\alpha\beta}\,
{\rm Tr}\left(F_{\nu\alpha}F_{\beta z}\right)\quad.\label{holocurrent}
\ee
Upon expanding $A_M$ in terms of the group field $U(x)$ (as well as excited mesons)
and performing $z$-integration,
we naturally expect
that it reduces to the usual Skyrmion number current in (\ref{skyrmion}).

A crucial point is that the above baryon current (\ref{holocurrent})
remains gauge invariant even in the presence of  background gauge potentials
for the chiral symmetry, which are encoded as non-normalizable modes of $A_M$.
Its conservation, relying on the Bianchi identity, is also intact
in any smooth situations up the boundary term in (\ref{bdr}). However, this boundary term
cancels in (\ref{consv}) in the case of vector-like background field, that is,
the fields coupled to $U(N_F)_I$ such that $A_\mu(+\infty)=A_\mu(-\infty)$.
The electromagnetism belongs to this case with
\be
A(+\infty)=A(-\infty)=Q A^{EM}\quad,\quad
Q=\left(\begin{array}{cc}{2\over 3}
& 0\\0 & -{1\over 3}\end{array}\right)\quad.\label{elec}
\ee
Because these two constraints uniquely fix
the baryon current (\ref{newbn}) in the presence of background electromagnetic potential,
the above holographic baryon current must reproduce (\ref{newbn})
as its lowest component involving $U(x)$.

To check this, it is convenient to work in the $A_z=0$ gauge
with an expansion~\cite{Sakai:2004cn}
\be
A_\mu(x,z)=A_{L\mu}^{\xi_+}(x)\psi_+(z)+A_{R\mu}^{\xi_-}(x)\psi_-(z)+({\rm excited\,\,modes})\quad,
\ee
where
\be
A_{L\mu}^{\xi_+}=\xi_+ (A_L)_\mu\xi_+^{-1} +\xi_+\partial_\mu\xi_+^{-1}\quad,\quad
A_{R\mu}^{\xi_-}=\xi_- (A_R)_\mu\xi_-^{-1} +\xi_-\partial_\mu\xi_-^{-1}\quad.
\ee
The group field $U(x)$ is contained in the above by $\xi^{-1}_+\xi_-=U$, and we denote the background
gauge potential by $A_L=A(+\infty)$ and $A_R=A(-\infty)$.  There still remains a residual gauge symmetry
to fix, called Hidden Local Symmetry, which is nothing but the gauge transformation at the
deepest IR $z=0$ which acts on the partial Wilson lines $\xi_\pm$ as $\xi_\pm(x)\to h(x)\xi_\pm(x)$.
For our purpose, we take the gauge $\xi_+^{-1}=U$ and $\xi_-=1$, upon which we have
\be
A_\mu=\left[\left(U^{-1}QU\right)\psi_+ + Q\psi_-\right]A_\mu^{EM} + \psi_+U^{-1}\partial_\mu U
+({\rm excited\,\,modes})\quad,
\ee
where we have explicitly used the electromagnetic background potential (\ref{elec}).
The details of the zero mode wavefunctions $\psi_\pm(z)$ won't be important later,
except $\psi_+ + \psi_-\equiv 1$ and $\psi_+(\infty)=\psi_-(-\infty)=1$.
From
\bear
F_{\nu\alpha}&=&\left(\left(U^{-1}QU-Q\right)\psi_+ +Q\right) F_{\nu\alpha}^{EM}
-\psi_+(1-\psi_+)\left[U^{-1}\partial_\nu U, U^{-1}\partial_\alpha U\right]\nonumber\\
&& +\psi_+(1-\psi_+)\left(U^{-1}\partial_\nu U\left(Q-U^{-1}QU\right)+\left[U^{-1},Q\right]
\partial_\nu U\right) A_\alpha^{EM} -(\nu\leftrightarrow\alpha)\quad,\nonumber\\
F_{\beta z}&=&-\left(\partial_z \psi_+\right)\left(\left(U^{-1}QU-Q\right) A_\beta^{EM}
+U^{-1}\partial_\beta U \right)\quad,
\eear
and integrating over $z$ in (\ref{holocurrent}), we can easily check that
the result indeed agrees with the previously known 4D result (\ref{newbn}).
Observe that the $z$-integration involves only the following integrals
\be
\int_{-\infty}^{+\infty}dz\, (\partial_z\psi_+)(\psi_+)^n ={1\over n+1}(\psi_+)^{n+1}\Big|^{+\infty}_{-\infty}
={1\over n+1}\quad,
\ee
without regard to a detailed functional form of $\psi_+$. This can be understood because
the final result is dictated by symmetry and it should be true
universally for any holographic model of QCD.

In the presence of a general chiral background potential, we naturally propose
(\ref{holocurrent}) to be the right answer for the modified baryon current.
With this granted, a violation of the baryon number by the boundary term (\ref{bdr})
\be
\partial_\mu B^\mu \sim \epsilon^{\mu\nu\alpha\beta}{\rm Tr}
\left(F_{\mu\nu}F_{\alpha\beta}\right)\Big|_R-\epsilon^{\mu\nu\alpha\beta}{\rm Tr}
\left(F_{\mu\nu}F_{\alpha\beta}\right)\Big|_L\quad,
\ee
implies that baryon number can be generated in
an environment with non-zero chiral-asymmetric sphaleron density.
This can be achieved by sphalerons made of elecro-weak gauge bosons, which
are indeed known to induce baryon asymmetry via chiral anomaly. Our result
can be thought of as
 a manifestation of this physics
in the low energy effective theory.

For a reference, we obtain from (\ref{holocurrent}) the baryon
current in a general chiral background potential $A_L$ and $A_R$,
\begin{eqnarray}
B^\mu&=&\frac{1}{24\pi^2}\,\epsilon^{\mu\nu\alpha\beta}\,{\rm Tr}\,
\left(U^{-1}\partial_{\nu}UU^{-1}\partial_{\alpha}UU^{-1}\partial_{\beta}U \right)\nonumber\\
& & -\frac{1}{8\pi^2}\,\epsilon^{\mu\nu\alpha\beta} {\rm
Tr}\,\partial_{\nu}
\left(U^{-1}A_{L\alpha}\partial_{\beta}U+A_{R\alpha}U^{-1}\partial_{\beta}U-
U^{-1}A_{L\alpha}UA_{R\beta} \right)\nonumber\\
& & -\frac{1}{8\pi^2}\,\epsilon^{\mu\nu\alpha\beta}\,{\rm Tr}\left(
\partial_{\nu}A_{L\alpha}\,A_{L\beta}+\frac23A_{L\nu}A_{L\alpha}A_{L\beta}-(L\leftrightarrow R)\right)\,.
\end{eqnarray}
This is an extension of (\ref{newbn}), which we find via holographic
QCD.

\section{Holographic Monopole Catalysis of Baryon Decay}

In this section, we re-analyze the monopole-induced baryon decay we discussed in
section \ref{section2} in the framework of holographic QCD, and find that the physics
becomes more transparent in holographic QCD.

A 4D background electromagnetic Dirac monopole enters our holographic model
as a specific non-normalizable mode in the expansion of 5D gauge field $A_M$.
In the gauge where we keep $A_z$ free, which is more convenient than the previous $A_z=0$ gauge
for the present purpose,
the background potential appears in the expansion as
\be
A_\mu(x,z)={1\over 2}\left(A_L +A_R\right)_\mu +
{1\over 2}\left(A_L -A_R\right)_\mu\psi_0(z)+({\rm normalizable\,\,modes})\label{exp}
\quad,
\ee
where $\psi_0(z)= {2\over \pi}\tan^{-1}(z)$ with $\psi_0(+\infty)=-\psi_0(-\infty)=1$.
$A_z$ contains only normalizable modes. For electromagnetism which is vector-like,
$A_L=A_R=QA^{EM}$, the second term is absent and we observe that a 4D monopole background
would enter the 5D expansion homogeneously along the $z$-direction.
In fact, a 5D gauge theory doesn't allow topological monopoles, but instead
can have  string-like objects (monopole-string) whose 3-dimensional transverse profiles
resemble those of monopoles. Therefore, the natural holographic object corresponding
to a 4D monopole is a monopole-string extending along the radial direction.
Behaviors of normalizable modes, including the pions $U(x)$, can be analyzed
by studying the 5D gauge field fluctuations around the monopole-string background.

A nice thing in this holographic set-up is that the violation of baryon number (\ref{holocurrent})
in the presence of a monopole-string has a simple explanation in terms of
a violation of the Bianchi identity due to the magnetic source.
The basic reason behind this simplification is that holographic QCD
unifies dynamical degrees of freedom of the model, such as $U(x)$,
with the background potential $A^{EM}$
in a single 5D gauge theory framework, so that their physics should find its explanations within
the 5D gauge theory.
We should also point out that in the monopole-string background, the normalizable modes
in the above expansion (\ref{exp}) will in general be excited by back-reactions, and the full
field configuration is more complicated than the Dirac monopole alone.
However, the amount of violation of the Bianchi identity
doesn't depend on these meson clouds due to its topological nature,
and is localized at the core of the monopole-string.

We write the 5D gauge field as
\be
A=Q A^{EM} + \tilde A\quad,
\ee
where $A^{EM}$ is a unit monopole-string background homogeneous along $z$ with $A^{EM}_z=0$,
and $\tilde A$ encodes any smooth dynamics of normalizable modes including pions $U(x)$, as well
as additional smooth background potential $A_L$ and $A_R$ coupled to the chiral currents.
Being a unit magnetic source, $A^{EM}$ is characterized by
\be
-2\pi i =\int_{S^2}\,F^{EM}=\int_{S^2}\,dA^{EM}=\int_{B^3}\,d^2 A^{EM}\quad,
\ee
where in the last equality we use the Stokes theorem on the 3-ball $B^3$ around the monopole core.
This gives us
\be
d^2 A^{EM}=-2\pi i \,\delta_3(\vec 0)\quad,
\ee
where $\delta_3(\vec 0)$ is a delta 3-form localized in space $\vec x$ at the monopole center $\vec x=\vec 0$.
In components $\delta_3(\vec 0)=\delta^{(3)}(\vec x)\,dx^1\wedge dx^2 \wedge dx^3$.
With this gadget, it is straightforward to find the Bianchi identity violation in 5D,
\be
DF\equiv dF+A\wedge F-F\wedge A = Q \,d^2 A^{EM}=-2\pi i \,Q\,\delta_3(\vec 0)\quad,
\ee
so that ${\rm Tr}\left(F\wedge F\right)$ is no longer closed,
\be
d\,{\rm Tr}\left(F\wedge F\right)=2{\rm Tr}\left(DF\wedge F\right)=
-4\pi i \,{\rm Tr}\left(QF\right)\wedge \delta_3(\vec 0)\quad.
\ee
In components this is equivalent to
\be
{1\over 4}\epsilon^{MNPQR}\partial_M{\rm Tr}\left(F_{NP}F_{QR}\right)
=4\pi i \delta^{(3)}(\vec x){\rm Tr}\left(Q F_{tz}\right)\quad,
\ee
which implies that
\be
\partial_\mu\left(\epsilon^{\mu\nu\alpha\beta}{\rm Tr}\left(F_{\nu\alpha}F_{\beta z}\right)\right)
=-{1\over 4}\partial_z\left(\epsilon^{\mu\nu\alpha\beta}
{\rm Tr}\left(F_{\mu\nu}F_{\alpha\beta }\right)\right)
+4\pi i \delta^{(3)}(\vec x){\rm Tr}\left(Q F_{tz}\right)\quad.
\ee
This is precisely what we need to find the violation of baryon number (\ref{holocurrent}),
\bear
\partial_\mu B^\mu&=&{1\over 8\pi^2}\int_{-\infty}^{+\infty} dz\,
\partial_\mu\left(\epsilon^{\mu\nu\alpha\beta}{\rm Tr}\left(F_{\nu\alpha}F_{\beta z}\right)\right)\\
&=&
{1\over 32\pi^2}\left(\epsilon^{\mu\nu\alpha\beta}{\rm Tr}
\left(F_{\mu\nu}F_{\alpha\beta}\right)\Big|_R-\epsilon^{\mu\nu\alpha\beta}{\rm Tr}
\left(F_{\mu\nu}F_{\alpha\beta}\right)\Big|_L\right)+{i\delta^{(3)}(\vec x)\over 2\pi}
\int_{-\infty}^{+\infty}dz\,{\rm Tr}\left(Q F_{tz}\right),\nonumber
\eear
where we will ignore the first term as we already discuss it in the previous section.

To study the monopole-induced second term,
let us go back to the $A_z=0$ gauge and expand $A_\mu(x,z)$ more precisely,
\be
A_\mu=\left[\left(U^{-1}(QA_\mu^{EM}+A_{L\mu} )U\right)\psi_+ + (QA_\mu^{EM}+A_{R\mu})\psi_-\right]
+\psi_+U^{-1}\partial_\mu U+\sum_{k\ge 1} B^{(k)}_\mu \psi_k\quad,
\ee
including now the complete spectrum of excited (axial) vector mesons in the expansion.
Since $A_t^{EM}=0$, and $F_{tz}=-\partial_z A_t$ in our gauge, the $z$-integral is readily performed
to give
\bear
\partial_\mu B^\mu&=&-{i\delta^{(3)}(\vec x)\over 2\pi}{\rm Tr}\left(Q A_t\right)\Big|^{+\infty}_{-\infty}
\nonumber\\
&=&-{i\delta^{(3)}(\vec x)\over 2\pi}\left[{\rm Tr}\left(Q U^{-1}\partial_t U\right)
+{\rm Tr}\left(Q U^{-1}A_{Lt} U\right)-{\rm Tr}\left(Q A_{Rt}\right)\right]\quad.\label{chemical}
\eear
This is the main result in this section.
Note that there is no contribution from excited (axial) vector mesons because
their wavefunctions $\psi_k(z)$ vanish sufficiently fast near the boundaries.
It is easy to see that the first term precisely reproduces to the 4D result (\ref{change}).
Writing $U(x)={\rm exp}({2i\over F_\pi} \pi^0(x) \sigma^3)$ as before, we have
\be
\partial_\mu B^\mu=
-{i\delta^{(3)}(\vec x)\over 2\pi}{\rm
Tr}\left(Q\sigma^3\right){2i(\partial_t \pi^0)\over F_\pi}=
{(\partial_t\pi^0)\over \pi F_\pi} \delta^{(3)}(\vec x)\quad,
\ee
whose $\vec x$-space integration is nothing but (\ref{change}).

It is also interesting
to speculate the implication of the second and the third terms. They seem to
indicate that in the presence of a chiral asymmetric chemical potential,
monopoles can create/annihilate baryon number. It would be interesting
to understand this better.

\section{String Theory Realization}

There is a nice stringy set-up realizing the physics of the previous sections
in terms of $D$-~branes
in the Sakai-Sugimoto model.
Let us parameterize the cigar-shaped part of the gravity background by
a radial coordinate $U\ge U_{KK}$ and an angle $\tau\sim\tau+2\pi M_{KK}^{-1}$~\cite{Sakai:2004cn}.
Details of the parameters $U_{KK}$ and $M_{KK}$ are not relevant in our discussion.
Our $N_F$ probe $D8$ branes are
spanning a line $\{\tau=0\} \cup \{\tau=\pi M_{KK}^{-1}\}$, $U\ge U_{KK}$ in the cigar part.
They also wrap the internal $S^4$ fibration and span the Minkowski space $R^{1,3}$.
We then consider a $D6$ brane which wraps the internal $S^4$ fibration
and spans a half of the cigar $0 \le \tau \le \pi M_{KK}^{-1}$, $U\ge U_{KK}$,
ending on one of the $N_F$ $D8$ branes. It is point-like in
the spatial $\vec x$ and static along the time.

Ignoring $S^4$ and $U$ directions
since they are common to the $D8$ and $D6$ branes, the system
is similar to a $D1$ brane ending perpendicularly on one of $N_F$ $D3$ branes.
The end point of $D1$ on $D3$ looks like a monopole source in view of $D3$ world volume
gauge field.
Because only one end of the $D1$ ($D6$) brane is on the $D3$ ($D8$) brane while the
other end extends to infinity,
the resulting monopole configuration on the $D3$ ($D8$) brane world volume
is an Abelian Dirac monopole with infinite energy in the spatial $\vec x$ directions.
Including the radial direction $U$ in the 5D $D8$ brane world volume (we ignore $S^4$),
or the $z$ direction in the previous sections,
what we have is precisely a 5D holographic, Dirac monopole-string
in the 5D gauge theory on the $D8$ branes.
Its $U(1)$ monopole charge direction depends on which $D8$ brane the $D6$ brane ends,
and we simply call it the charge matrix $Q$.
For an example of $N_F=2$ with the $D6$ brane ending on the first $D8$ brane,
we have
\be
Q=\left(\begin{array}{cc} 1& 0 \\ 0& 0\end{array}\right)\quad.
\ee
As the monopole-string is homogeneous along $U$ (or $z$), it represents
a vector-like background gauge potential of the chiral symmetry (or $A_L=A_R=Q A^{Dirac}$)
with a monopole charge $Q$ in holographic QCD.
Therefore, the physics of monopole catalysis of baryon decay
that we study in the previous sections must apply to this stringy setting.

Indeed, we can easily identify the string theory
phenomenon corresponding to the baryon decay in the presence of a $D6$ monopole-string.
5D instanton-baryons on the $D8$ branes can be thought of as $S^4$-wrapped
$D4$ branes dissolved into the $D8$ branes.
But, these $S^4$-wrapped $D4$ branes can also dissolve into our $D6$ brane, because
they are similar to a $D0/D2$ system when we ignore the common $S^4$ directions.
Therefore, $D4$-baryons can be captured by the $D6$ monopole-string and disappear from
the $D8$ world volume. This is the string theory correspondent to the monopole catalysis
of baryon decay.

We define the baryon number of a single $S^4$-wrapped $D4$ brane to
be one, as it also has unit instanton number on the $D8$ branes. The
dissolved $D4$-baryon number into our $D6$ monopole-string is
measured by the 2-form field strength $F^{(2)}=dA^{D6}$ of the $D6$
world volume gauge potential on the half cigar that the $D6$ world
volume spans, similar to $D0/D2$ system, \be (\Delta B)_{D6}={i\over
2\pi}\int_{(U,\tau)}\,F^{(2)} =-{i\over 2\pi}
\int_{-\infty}^{+\infty} dz\, A_z^{D6}\quad, \ee where the
normalization can be fixed by that a unit $(-2\pi i)$ flux
represents a single dissolved $D4$-brane, and we use the Stokes
theorem in the last equality since the boundary of the $D6$ half
cigar is precisely the line along $z$ at the $D6/D8$ intersection.
Note that in $\vec x$ space, this is precisely the position of the
monopole-string core. As the $D6$ brane ends on one of the $D8$
branes, the $D6$ world volume gauge potential is identical to the
corresponding $D8$ world volume gauge potential at the $D6/D8$
intersection. Therefore, the above $A_z^{D6}$ can be equally
interpreted as the $A_z$ at the monopole-string core on the $D8$
brane on which the $D6$ brane ends, \be A_z(t,\vec 0) = Q
A_z^{D6}(t)\quad, \ee where we take the monopole position at the
origin $\vec x=\vec 0$, and $Q$ represents
the charge matrix of $D6$ ending on the $D8$ brane, as given before. Using
${\rm Tr}(Q^2)=1$, we have \be (\Delta B)_{D6}=-{i\over
2\pi}\int_{-\infty}^{+\infty} dz\,{\rm Tr}\left(Q
A_z\right)\Big|_{\vec x=\vec 0}\quad, \ee whose time derivative must
be equal to the rate of disappearance of the baryon number from the
$D8$ branes, \be {dB\over dt} = -{d \over dt}(\Delta B)_{D6}
=-{i\over 2\pi}{d \over dt}{\rm Tr}\left( -Q\int_{-\infty}^{+\infty}
dz\,A_z\right)\quad. \ee Noting that the integral inside the trace
is precisely the Wilson line corresponding to the pions, \be
-\int_{-\infty}^{+\infty} dz\,A_z= {2i \over F_\pi}\pi^0 (\vec
0)\sigma^3 \quad, \ee we finally have \be {dB\over dt} ={1\over \pi
F_\pi}{\rm Tr}\left(Q\sigma^3\right)\left(\partial_t
\pi^0\right)\quad. \ee This is precisely what we have in the
previous sections.

\section{Conclusion}

Fermion number is often not conserved in background fields which
modify the spectrum of fermions~\cite{Rubakov:2002fi}. One
tantalizing such phenomenon is the monopole catalysis of baryon
decay, where baryons disappear (or appear) near the magnetic
monopole. Since monopole catalysis of baryon decay may have a
significant effect in monopole search and proton decay experiment,
both of which are consequences of unified gauge theories, it is
desirable to understand it more clearly. We have investigated
monopole catalysis in the context of the gauge/string duality and
showed how it is realized in holographic QCD and also in string
theory. In doing so we have demonstrated that the gauge/string
duality is indeed a powerful tool to study strong interactions, and
found that the baryon number violation under the magnetic monopole
or by the electroweak sphaleron can be formulated into a single
equation in holographic QCD.

There are several phenomenological implications of our study. One of
them  is the generation of baryons in the presence of magnetic
monopole by external chiral chemical potentials, as shown in
Eq.~(\ref{chemical}), which might be more effective in generating
baryon asymmetry at lower temperature where the sphalerons are
suppressed. This mechanism might be relevant in early universe or
heavy ion collision but we leave it for future investigation.

\subsection*{Acknowledgments}
This work is supported in part
by the Korea Research Foundation Grant funded by the
Korean Government (MOEHRD, Basic Research Promotion Fund)
(KRF-2007-314-C00052)(D.~K.~H.),
by KOSEF
Basic Research Program with the grant No. R01-2006-000-10912-0 (D.~K.~H. and C.~P.),
by the KOSEF SRC Program through CQUeST at Sogang University (K.~M.~L.),
KRF Grants No. KRF-2005-070-C00030 (K.~M.~L.), and
the KRF National Scholar program (K.~M.~L.).
H.U.Y. thanks Koji Hashimoto, Takayuki Hirayama and Feng-Li Lin for discussions.

%\vfill
%\eject

\end{document}